\documentclass{ws-ijmpa}

\begin{document}

\markboth{M. {\L}uszczak and A. Szczurek}
{Charmed meson production}

%
\catchline{}{}{}{}{}
%

\title{CHARMED MESON PRODUCTION IN PROTON - (ANTI)PROTON COLLISIONS
\footnote{Based on a talk given by M. {\L}uszczak at MESON2006
(Krak\'ow, June 9-13, 2006)}
}

\author{\footnotesize MARTA {\L}USZCZAK}

\address{University of Rzesz\'ow, ul. Rejtana 16c\\
PL-35-959 Rzesz\'ow, Poland
}

\author{ANTONI SZCZUREK}

\address{Institute of Nuclear Physics PAN, ul. Radzikowskiego 152\\
PL-31-342 Cracow, Poland\\
and \\
University of Rzesz\'ow, ul. Rejtana 16c\\
PL-35-959 Rzesz\'ow, Poland
}

\maketitle

\pub{Received (Day Month Year)}{Revised (Day Month Year)}

\begin{abstract}
We discuss and compare different approaches to include gluon 
transverse momenta for heavy quark-antiquark pair and meson production.
The results are illustrated with the help
of different unintegrated gluon distributions (UGDF)
from the literature. We compare results obtained with
on-shell and off-shell matrix elements and kinematics.
The results are compared with recent experimental results of
the CDF collaboration.

\end{abstract}

\section{Introduction}

The heavy quark-antiquark production in hadroproduction
is known as one of the crucial tests of conventional 
gluon distributions within a standard factorization approach.
At high energies one tests gluon distributions at low values
of longitudinal momentum fraction.
Standard collinear approach does not include transverse momenta of initial gluons, the method to include transverse momenta is $k_t$ - factorization approach\cite{CCH91,CE91,BE01}.
In the first step of our approach the single
particle spectra of charmed quarks and antiquarks are obtained assuming
gluon-gluon fusion\cite{LS06}. Different unintegrated gluon distributions from the literature are used\cite{KL01,AKMS94,GBW_glue,KMR,BFKL}. To obtain the single particle spectra of meson
from those of quarks/antiquarks a standard hadronization procedure
with Peterson fragmentation function is applied.
Conclusions about unintegrated gluon distributions are drawn.

\section{Heavy quark production}



\begin{figure}[!htb] 
\begin{center}
\includegraphics[width=3.5cm]{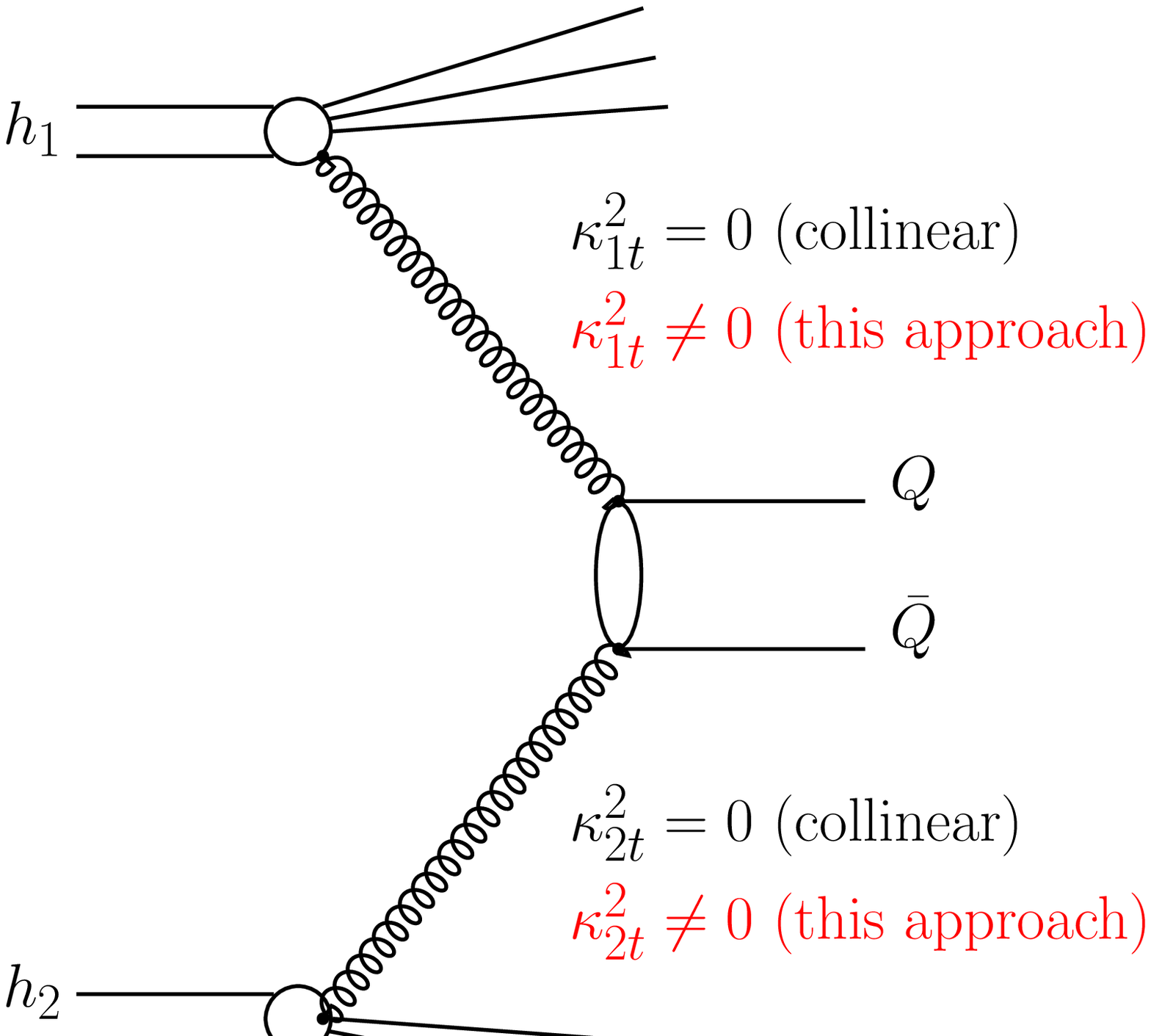}
\includegraphics[width=3cm]{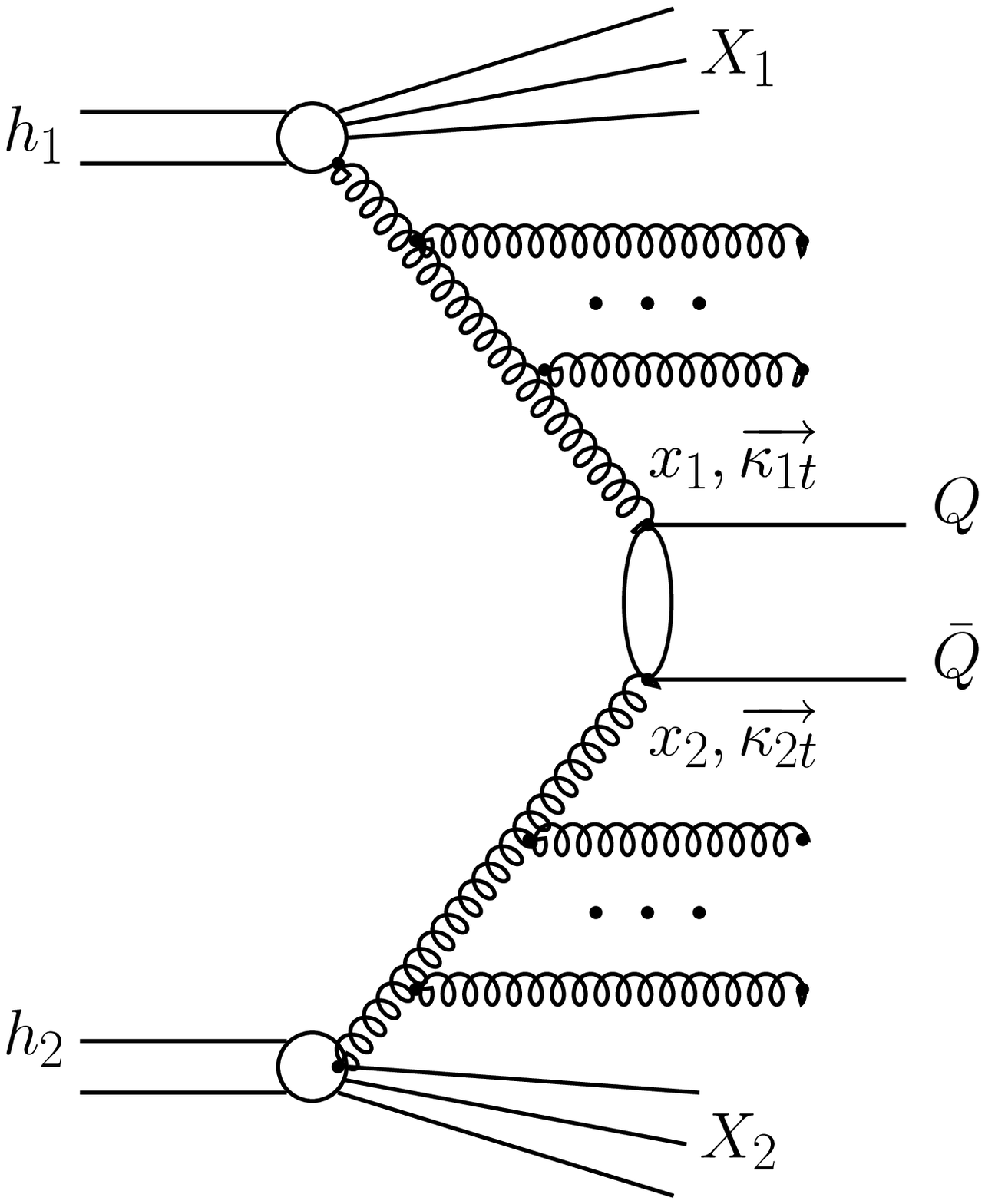}
\caption[*]{
Leading-order diagrams for heavy quark - antiquark production.
\label{fig:d_gluon}
}
\end{center}
\end{figure}


Let us consider the reaction $h_1 + h_2 \to Q + \bar Q + X$,
where $Q$ and $\bar Q$ are heavy quark and heavy antiquark,
respectively.

In the leading-order approximation within collinear approach
the triple-differential cross section in rapidity of $Q$ ($y_1$),
in rapidity of $\bar Q$ ($y_2$) and transverse momentum of
one of them ($p_t$) can be written as
\begin{equation}
\frac{d \sigma}{d y_1 d y_2 d^2p_t} = \frac{1}{16 \pi^2 {\hat s}^2}
\sum_{i,j} x_1 p_i(x_1,\mu^2) \; x_2 p_j(x_2,\mu^2) \;
\overline{|{\cal M}_{ij}|^2} \; .
\label{LO_collinear}
\end{equation}
Above $p_i(x_1,\mu^2)$ and $p_j(x_2,\mu^2)$ are familiar
(integrated) parton distributions in hadron $h_1$ and $h_2$, respectively.

 The parton distributions are evaluated at:
$x_1 = \frac{m_t}{\sqrt{s}}\left( \exp( y_1) + \exp( y_2) \right)$,
$x_2 = \frac{m_t}{\sqrt{s}}\left( \exp(-y_1) + \exp(-y_2) \right)$.
The formulae for matrix element squared averaged over initial
and summed over final spin polarizations can be found e.g. in
Ref.\cite{BP_book}.

If one allows for transverse momenta of the initial partons,
the transverse momenta of the final $Q$ and $\bar Q$ no longer
cancel.
Formula (\ref{LO_collinear}) can be easily generalized if
one allows for the initial parton transverse momenta. Then
\begin{eqnarray}
\frac{d \sigma}{d y_1 d y_2 d^2p_{1,t} d^2p_{2,t}} = \sum_{i,j} \;
\int \frac{d^2 \kappa_{1,t}}{\pi} \frac{d^2 \kappa_{2,t}}{\pi}
\frac{1}{16 \pi^2 (x_1 x_2 s)^2} \; \overline{ | {\cal M}_{ij} |^2}
\nonumber \\  
\delta^{2} \left( \vec{\kappa}_{1,t} + \vec{\kappa}_{2,t} 
                 - \vec{p}_{1,t} - \vec{p}_{2,t} \right) \;
f_i(x_1,\kappa_{1,t}^2) \; f_j(x_2,\kappa_{2,t}^2) \; ,
\label{LO_kt-factorization}    
\end{eqnarray}
where now $f_i(x_1,\kappa_{1,t}^2)$ and $f_j(x_2,\kappa_{2,t}^2)$
are so-called unintegrated parton distributions.
The extra integration is over transverse momenta of the initial
partons.
The two extra factors $1/\pi$ attached to the integration over
$d^2 \kappa_{1,t}$ and $d^2 \kappa_{2,t}$ instead over
$d \kappa_{1,t}^2$ and $d \kappa_{2,t}^2$ as in the conventional
relation between unintegrated and integrated parton distributions. 
The two-dimensional delta function assures momentum conservation.
Now the unintegrated parton distributions must be evaluated at:
$x_1 = \frac{m_{1,t}}{\sqrt{s}} \exp( y_1) +
       \frac{m_{2,t}}{\sqrt{s}} \exp( y_2)$,
$x_2 = \frac{m_{1,t}}{\sqrt{s}} \exp(-y_1) +
       \frac{m_{2,t}}{\sqrt{s}} \exp(-y_2)$.
In general, the matrix element must be calculated for initial
off-shell partons. The corresponding formulae for initial gluons
were calculated in \cite{CCH91,CE91} (see also \cite{BE01}).
In the present paper for illustration we shall compare results
obtained for both on-shell and off-shell matrix elements.


\begin{figure}[!thb] 
\begin{center}
\includegraphics[width=6cm]{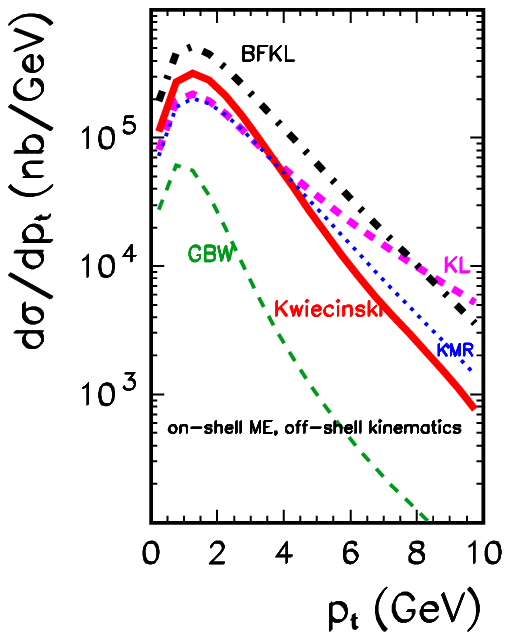}
\includegraphics[width=6cm]{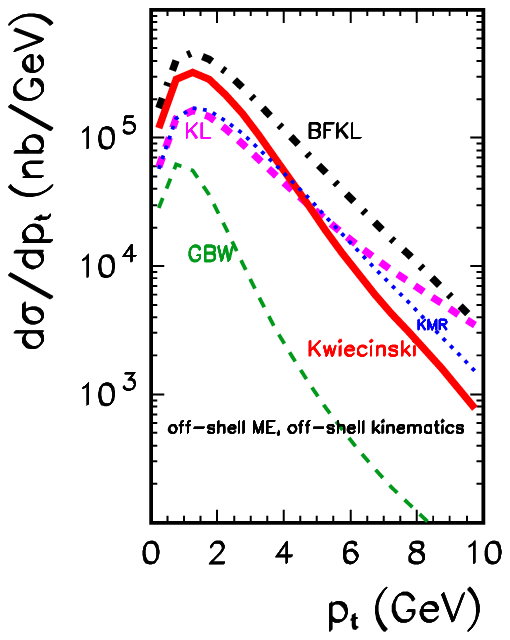}
\caption[*]{
Inclusive $d\sigma/d p_t$ for charm/anticharm production at W = 1.96 TeV
for different UGDF.
The results with on-shell kinematics are shown in panel
(a) and results with off-shell kinematics in panel (b).
In this calculation both factorization and renormalization scales
were fixed for 4 $m_c^2$.

\label{fig:dsig_dpt_gauss}
}
\end{center}
\end{figure}

In Fig.2  we collected results for $d \sigma/ dp_t$ obtained with
different unintegrated gluon distributions from the literature.
In this case consequently the off-shell matrix element and
off-shell kinematics were used. The GBW gluon distribution leads to a
much smaller cross section. The KL gluon distribution produces
the hardest $p_t$ spectrum. Rather different slopes in
transverse momentum of c (or $ \bar c$) are obtained for different
UGDFs.
This differences survive after convoluting the inclusive
quark/antiquark spectra with fragmentation functions. Thus, in
principle, precise distribution in transverse momentum of charmed
mesons should be useful to select a ''correct'' model of UGDF.
A detailed comparision with the experimental data requires, however,
a detailed knowledge of fragmentation functions.

The inclusive spectra are not the best observables to test
UGDF \cite{LS04}.
Let us come now to correlations between charm quark and
charm antiquark.


\begin{figure}[!thb] 
\begin{center}
\includegraphics[width=6.0cm]{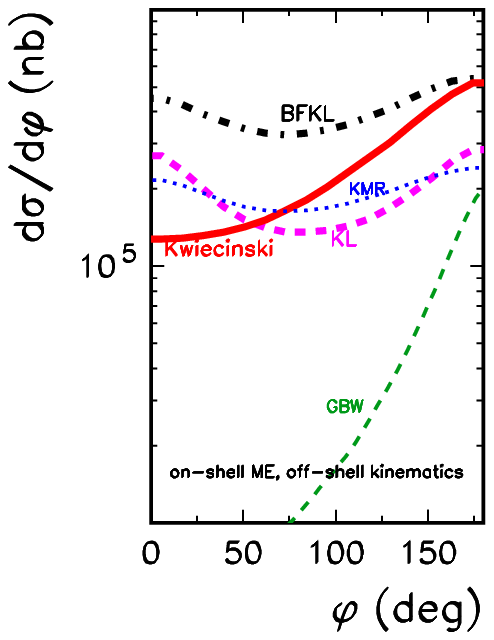}
\includegraphics[width=6.0cm]{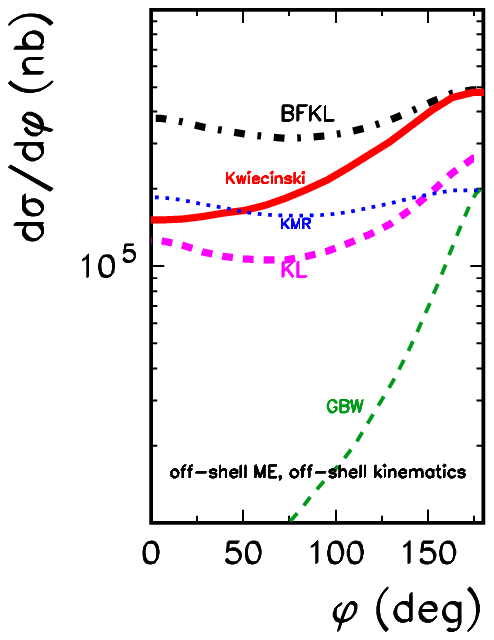}
\caption[*]{
Azimuthal angle correlations for different UGDFs in the literature:
Kwieci\'nski (solid), BFKL (dash-dotted), GBW (dashed) and KL (dotted).
\label{fig:dsig_dpphi_ugdf}
}
\end{center}
\end{figure}


In Fig.3 we compare results for different
unintegrated gluon distribution from the literature.
Quite different results are obtained for different UGDFs.
The nonperturbative GBW glue leads to strong azimuthal correlations
between $c$ and $\bar c$. In contrast, BFKL dynamics leads to strong
decorrelations of azimuthal angles of charm and anticharm quarks.
The saturation idea inspired KL distribution leads to an local
enhancement for $\phi_{c \bar c}$ which is probably due
to simplifications made in parametrizing the UGDF.
In the last case there is a sizeable difference between the result
obtained with on-shell (left panel) and off-shell (right panel)
matrix elements. All this is due to an interplay of the matrix element
and the unintegrated gluon distributions.

\section{From unintegrated parton distributions to meson
  production}

The inclusive distributions of hadrons
are obtained through a convolution of inclusive distributions
of heavy quarks/antiquarks and Q $\to$ h fragmentation functions
\begin{equation}
\frac{d \sigma(y_h,p_{t,h})}{d y_h d^2 p_{t,h}} \approx
\int_0^1 \frac{dz}{z^2} D_{Q \to h}(z)
\frac{d \sigma_{g g \to Q}^{A}(y_Q,p_{t,Q})}{d y_Q d^2 p_{t,Q}}
\Bigg\vert_{y_Q = y_h \atop p_{t,Q} = p_{t,h}/z}
 \; . 
\label{Q_to_h}
\end{equation}
In the present paper we have used so-called Peterson fragmentation
functions with parameters from last issue of PDG (Particle Data Group).
We have neglected a possibility that charmed mesons are produced
from light quarks and/or gluons. This approximation should be better
for heavier quarks/mesons.

In the present analysis we show dependence of the total cross section
on transverse momenta for $D^{*+}$ production

\begin{equation}
\frac{d \sigma(p_t)}{dp_t} =
\int_{-1}^{1} dy \; \frac{d \sigma(y,p_t)}{dydp_t} \approx
2 \frac{d \sigma(y=0,p_t)}{dydp_t}   \; .
\label{LO_kt-factorization}
\end{equation}


\begin{figure}[!htb] 
\begin{center}
\includegraphics[width=6cm]{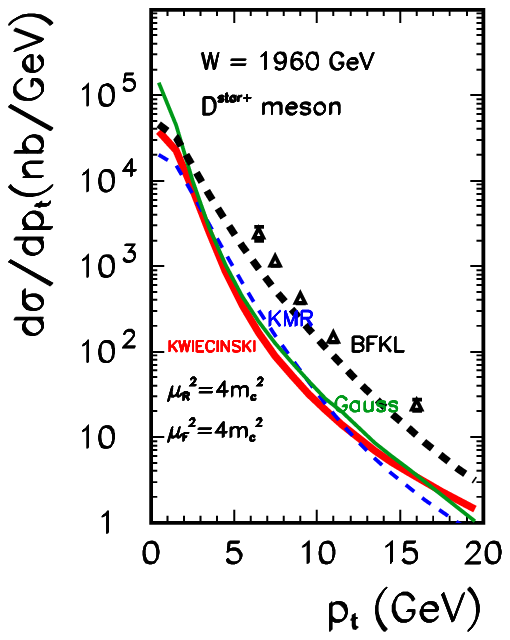}
\includegraphics[width=6cm]{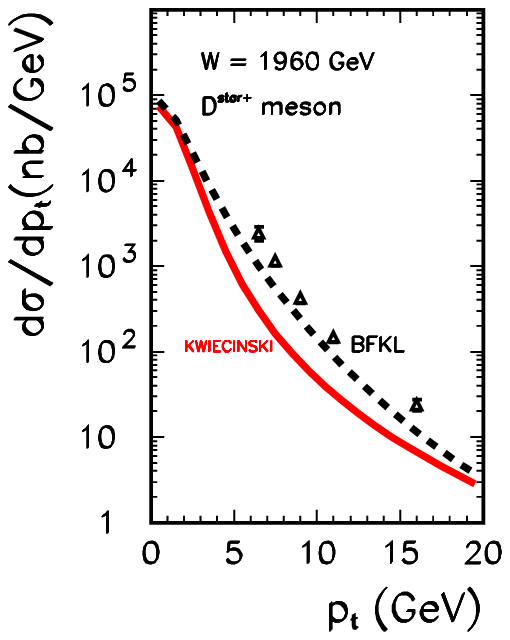}
\caption[*]{
Inclusive $d\sigma/d p_t$ for $D^{*+}$ production at W = 1.96 TeV
for different UGDF.
The results with $\alpha_s = \alpha_s (4m_c^2)$ is in the left panel
and the results with $\alpha_s = \alpha_s (\kappa_{1t}^2)$ or
$\alpha_s (\kappa_{2t}^2)$ in the right panel.
\label{fig:d_gluon}
}
\end{center}
\end{figure}


In Fig.4 the theoretical results are compared with experimental
data of the CDF collaboration \cite{D_CDF_data}. We collected results
obtained with different unintegrated gluon distributions from
the literature.

We have made calculations for two different choices of renormalization
scale. The results obtained with $\alpha_s(4 m_c^2)$ (left panel)
are below the experimental data, while results with
$\alpha_s = \alpha_s (\kappa_{1t}^2)$ or $\alpha_s (\kappa_{2t}^2)$
(see \cite{LSZ02}) better describe the data.
There is also a dependence on UGDF used in the calculation.
Even if the second choice is made, there seem to be a deficit
of the cross section.

It is not clear to us if the deficit is due to the omission of
$q \to D^*$ or $g \to D^*$ fragmentation, or due to higer order
terms in heavy quark production.

Similar situation is for $B^+$ production, as shown in Fig.5.


\begin{figure}[!htb] 
\begin{center}
\includegraphics[width=6cm]{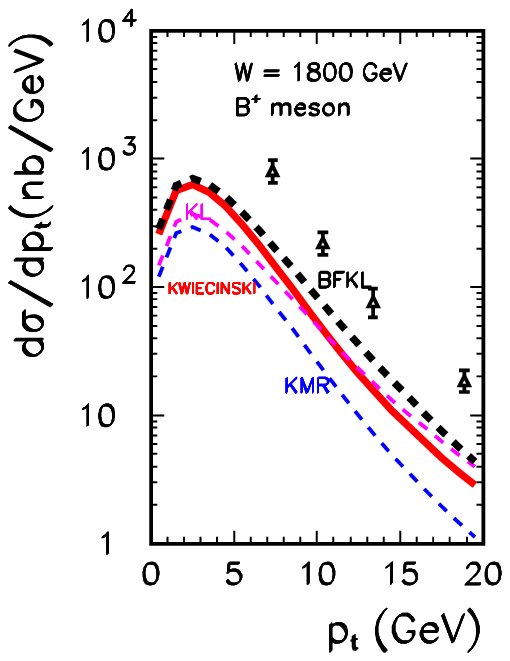}
\includegraphics[width=6cm]{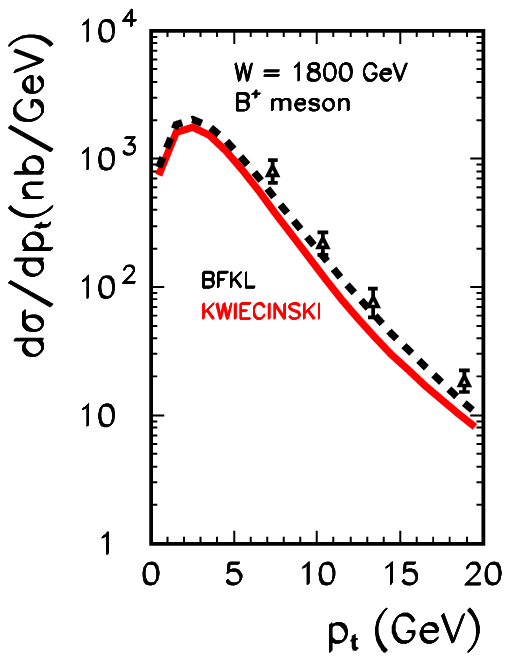}
\caption[*]{
Inclusive $d\sigma/d p_t$ for $B^+$ production at W = 1800 GeV
for different UGDF. 
The experimental data are from Ref.\cite{B_CDF_data}.
The results with $\alpha_s = \alpha_s (4m_c^2)$ is in the left panel
and the results with $\alpha_s = \alpha_s (\kappa_{1t}^2)$ or
$\alpha_s (\kappa_{2t}^2)$ in the right panel.
\label{fig:d_gluon}
}
\end{center}
\end{figure}


\section{Summary}

Inclusive cross section for heavy quark - antiquark and heavy mesons
in proton -(anti)proton collisions have been calculated 
in the $k_{t}$-factorization approach.
We have compared quantitatively different methods to include
gluon transverse momenta and their effect on inclusive spectra
as well as on $c \bar c$ correlations. Different UGDF from
the literature were used.

The inclusive spectra of heavy mesons were obtained via
convolution of heavy quark spectra with so-called Peterson fragmentation
functions. The results depend on the choice of the factorization
scale.
The ``best'' results are obtained with the BFKL UGDF.
We observe a deficit of the cross section for almost all UGDFs used.
It is not clear to us if the deficit is due to neglecting
some hadronization components or due to next-to-leading order
effects.



\end{document}